\def\etal{{et.al.}\/}
\def\H{{\bf H}\/}
\def\uul#1{\underline{\underline{#1}}\/}
\def\ul#1{\underline{#1}\/}

\def\n{\noindent}
\def\PRL{{\it Phys. Rev. Lett}}
\def\PR{{\it Phys. Rev. }}
\def\JPC{J. Phys. C }
\def\JPCM{J. Phys.: Condens. Matter }
\def\be{\begin{equation}}
\def\ee{\end{equation}}
\def\R{{\vec{R}}}

\documentclass{article}
\usepackage{graphicx}
\begin{document}
\title{Augmented space recursion code and application in simple binary metallic alloy}
\author{Rudra Banerjee and Abhijit Mookerjee\\Advanced Materials Research Unit, S N Bose National Centre for Basic
Sciences \\
Block JD, Sector III, Salt Lake, Kolkata 700098,India\\
rudra@bose.res.in}

%
%

\maketitle

\begin{abstract}
We present here an optimized and parallelized version of the {\sl augmented space recursion code} for the calculation of the electronic and magnetic
properties of bulk disordered alloys, surfaces and interfaces, either flat, corrugated or rough, and random networks. Applications have been made
to bulk disordered alloys to benchmark our code.
\end{abstract}

:
\section{Introduction}

The aim of this communication is to present and describe a computational package to handle first-principles density functional (DFT)
based studies of electronic structure of systems without long-ranged lattice translational symmetry. 
Bulk disordered alloys and surfaces and interfaces which are either flat, corrugated or rough fall under
this category which our formalism should be able to take care of. The aim is also to go beyond the usual mean-field
 approaches like the coherent potential approximation (CPA) and be able to take into account configuration
fluctuations of the local environment. Lack of lattice translational symmetry
means that the standard reciprocal space techniques based on the powerful Bloch theorem can no longer
be applicable  and we shall depend on alternative techniques based purely on real space approaches. Our
formalism will be a marriage of three distinct methods which have been individually applied extensively :
namely, the recursion method (RM) of Haydock \etal\ \cite{hhk}-\cite{hhk2}, the augmented space method proposed by one of us
\cite{asr1}-\cite{asr2} (ASR) and the tight-binding, linear muffin-tin orbitals method (TB-LMTO) \cite{ak1}. The last mentioned
provides us with a DFT self-consistent sparse representation of the Hamiltonian in a real-space minimal basis
$\{\vert \R_nL\rangle\}$ which spans the Hilbert space ${\cal H}$. Here $\R_n$ labels the sites
where   the ion-cores sit and $L=(\ell m\sigma)$ are the 
angular momentum indexes. For a disordered system the matrix elements of the Hamiltonian representation in this basis   are random.
This representation is then taken over by the ASR to generate a modified Hamiltonian representation 
in the outer product space of ${\cal H}$ and the space ${\cal C}$ of configuration fluctuations of the random parameters.
This modified Hamiltonian in the augmented space ${\cal H}\otimes{\cal C}$ represents a collection of all possible  Hamiltonians for all possible configurations
of Hamiltonian representations. Once this is done the RM  allows us to obtain the matrix elements of
the Green functions related to the Kohn-Sham equation for the electronic states. The augmented space theorem
(AST)\ \cite{ast} relates a specific matrix element of the Green function in the augmented space ${\cal H}\otimes
{\cal C}$ to the configuration averaged Green functions. The RM is a purely  real-space based technique,
and therefore as applicable to a bulk system as one with
surfaces, interfaces or extended defects. In the following we shall describe each of the points raised above
in some detail.

\section{Tight-binding linear muffin-tin orbitals method}

The TB-LMTO has been described in great detail earlier \cite{ak1},\cite{ak2}-\cite{ad}. We
shall only quote here the main results which will be relevant for setting up the ASR programme.
The starting point is the Kohn-Sham equation with the muffin-tin effective crystal potential.
The basis chosen for representation of the wave-function are  the  muffin-tin orbitals $\{\vert \R_nL\rangle\}$ :
If we  expand the wave-function  in terms of a linear combination of these 
muffin-tin orbitals and substitute the expansion in the Kohn-Sham equation we obtain the
Korringa-Kohn-Rostocker (KKR) secular equation~:

\[ \det \parallel P_{\R_nL}(E)\delta_{\R_n\R_m}\delta_{LL'} - S_{\R_nL,\R_nL'}(\kappa)
\parallel = 0 \]

Energy linearization gives the eigen-type LMTO secular equation:

\be  \det\parallel E\delta_{\R_n\R_m}\delta_{LL'} - H^{(2)}_{\R_nL,\R_mL'}\parallel = 0\ee

where, the `second order' Hamiltonian is given by :

\be
\uul{H}^{(2)}_{\R_n,\R_m}  =  \uul{E}_{\nu}\ \delta_{\R_n\R_m}+\uul{h}_{\R_n,\R_m}
 - \sum_{\R_n}\ \uul{h}_{\R_n,\R_k}\ \uul{o}_{\R_k}\  \uul{h}_{\R_k,\R_m}    
\label{ham}
\ee

Note that each element is a matrix in the $L$ space.
In the screened representation the structure matrix $\uul{S}_{\R_n\R_m}$ is short-ranged and 
the Hamiltonian is sparse~:

\begin{eqnarray}
\uul{h}_{\R_n,\R_m}  =  (\uul{C}_{\R_n} - \uul{E}_{\nu})\ \delta_{\R_n\R_m} + \uul{\Delta}^{1/2}_{\R_n}\ 
\uul{S}_{\R_n,\R_m}(0)\ \uul{\Delta}^{1/2}_{\R_m}
\end{eqnarray}

The expansion energies $E_{\nu L}$ are chosen suitably by us at the center of the energy window
of our interest and the potential parameters $\uul{C},\uul{\Delta}$ and $\uul{o}$ are diagonal
in $L$ space and are self-consistently
generated. The  structure matrix $\uul{S}$ is obtained from the geometry of the lattice.
Note that the TB-LMTO basis is minimal and hence usually it is enough to take $\ell \leq 3$
 and in a majority of cases one does not have to go beyond $\ell = 2$

At this point we should comment on several possible generalizations : we have energy linearized the secular equation. In case we do not wish to do so, we can still
deal with a energy dependent ``Hamiltonian" or ``secular matrix" $ \uul{R}(E)=\uul{P}(E)-\uul{S}(\kappa)$.
This is the TB-KKR. The subsequent recursion becomes energy-dependent, however each recursion at each energy point can 
 be parallely carried out for efficiency. Such energy dependent recursion has been carried out by us earlier \cite{enrec}. 
The assumption (a posteriori shown after calculations)  was that the recursion coefficients are weakly energy dependent,
therefore recursion is carried out an equi-spaced ``seed" points and the intermediate points found by interpolation. 
Further, if we allow for third-nearest neighbour sparsity in the Hamiltonian
we could start with the real-space full-potential LMTO (RS-FPLMTO) of Eriksson \etal \cite{fplmto}.

\section{The Recursion method}

Whereas,the above formulation is general enough, we should stress on the fact that the Hamiltonian
representation is an infinite matrix. Lack of lattice translation symmetry means that we are 
unable to symmetry reduce the dimensionality of the matrix to essentially $2\ell_{max}+1$.
We therefore have to resort to techniques which allow us to deal with infinite matrices. The Recursion
method was exactly such a technique introduced by Haydock \etal \cite{hhk}-\cite{hhk2} to obtain the Green
functions associated with the secular equation.  This would give us the density of states from which
we may obtain properties like the Fermi energy, the charge and magnetization densities, the magnetic
moment and the band energy. A recent generalization also gives us correlation functions associated with response
functions \cite{genrec1}-\cite{genrec2}. 

The RM  begins by recursively changing the basis through a three term recurrence relation :

\begin{eqnarray}
\vert 1\rangle & =  &  \vert \R_iL\rangle \qquad\mbox{and}\qquad 
\vert 2\rangle  =   \H^{(2)}\vert 1\rangle - \alpha_1 \vert 1\rangle \nonumber \\
\vert n+1\rangle & = &  \H^{(2)}\vert n\rangle - \alpha_n \vert n\rangle-\beta^2_n \vert n-1\rangle
\qquad\mbox{for}\quad n>1
\label{rec}
\end{eqnarray}

Mutual orthogonality of this new basis gives :

\begin{eqnarray}
 \alpha_n  = \frac{\langle n\vert \H^{(2)}\vert n\rangle}{\langle n\vert n\rangle}
\quad\mbox{and}\quad \beta_n^2  = \frac{\langle n+1\vert n+1\rangle}{\langle n\vert n\rangle}
\label{albe}
\end{eqnarray}

To understand how the above equations are efficiently operationally coded on the computer, let us first describe
the Hamiltonian as an operator \H$^{(2)}$. From Eqn.(\ref{ham})

\begin{eqnarray}
\H^{(2)} & = &  {\bf E}_\nu + {\bf h} - {\bf h}\ {\bf o}\ {\bf h} \quad\mbox{  and  }\quad 
 {\bf h}  =  {\bf C}-{\bf E}_\nu + {\mathbf \Delta}^{1/2} \ {\bf S}\ {\mathbf \Delta}^{1/2}
\label{ham2}
\end{eqnarray}

Of these operators four are `diagonal' : {\bf E}$_\nu$, {\bf C}, ${\mathbf \Delta}^{1/2}$ and {\bf o}.
Their structure are all of the form :
$ {\mathbf D} = \sum_i \ \uul{D}_i \ \vert R_i\rangle\langle R_i\vert = \sum_i \uul{D}_i{\cal{P}}_i$. While the structure matrix is off-diagonal :
 ${\mathbf S} = \sum_{(ij)} \uul{S}_{\R_i\R_j} \left(\vert\R_i\rangle\langle\R_j\vert
+\vert\R_j\rangle\langle\R_i\vert\right) = \sum_{(ij)} \uul{S}_{\R_i\R_j} {\cal{T}}_{ij}$. 

If we represent a Hilbert space `vector' $\vert n\rangle$ by a matrix as shown below~:

\[ \vert n\rangle \Longrightarrow \left( \begin{array}{cccc} 
                                      n_{11}& n_{12} & \ldots & n_{1L} \\
                                      n_{21}& n_{22} & \ldots & n_{2L} \\
                                            & \vdots &   \vdots &  \\
                                      n_{p1}& n_{p2} & \ldots & n_{pL} \\
                                            & \vdots &   \vdots &  \\
					\end{array}\right)  \Longrightarrow \left( \begin{array}{c}
										\ul{n}_1 \\
                                                                                \ul{n}_2 \\
										\vdots  \\
										\ul{n}_p\\
										\vdots \end{array}\right)
\]

The action of the diagonal operators {\bf D} is rather simple:
\begin{enumerate}
\item[(i)] One by one choose non-zero rows $\ul{n}_i$ of $\vert n\rangle$
\item[(ii)] Multiply $\uul{D}_i\ \ul{n}^T_i$ and add this to the $i$-th row of a new $\vert n'\rangle$.
\item[(iii)] Repeat this for all non-zero rows of $\vert n\rangle$. Finally $\vert n'\rangle = {\bf D}\vert n\rangle$
\end{enumerate}

\begin{figure}
\includegraphics[width=5in]{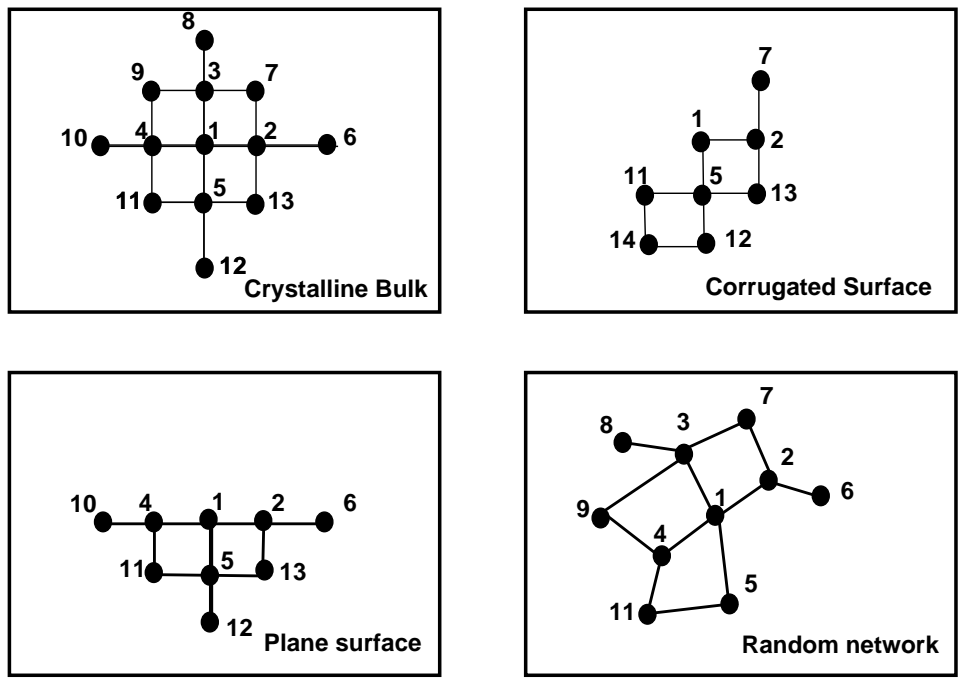}
\caption{The geometry of (top,left) a bulk square lattice (bottom,left) a square lattice with a
plane (100) surface (top,right) a square lattice with a corrugated (110) surface and (bottom,right)
a fourfold coordinated planar network} \label{map}
\end{figure}

The action of the off-diagonal operator {\bf S} is more complicated as it has the information of the
underlying lattice or network embedded in it. The first step would be to prepare the ``neighbour map" (NM).
The NM is a matrix $\uul{N}= N_{ij}$ where $N_{ij}$ is the $j$-th neighbour of the $i$-th site on the
lattice or network. We have first to number the lattice/network points by integers. The near neighbour
vectors point in different directions, while numbering this directionality has to be carefully
recorded, since the structure matrix between $\R_n$ and $\R_m$ will depend upon the direction of
$\R_n-\R_m$. In figure \ref{map} we show the geometry of a bulk square lattice, a square lattice with
a plane surface (100), one with a corrugated surface (110) and a network with four local bonds.

Since the integer numbering increases always clockwise from a site, the directionality is preserved.
The neighbour maps, also preserving directionality,  for the first three examples are~:

\[ \hskip -1cm \left(\begin{array}{cccc|c}
        2 \ &\  3 \ &\  4 \ &\  5 \ &\  4\\
        6 \ &\  7 \ &\  1 \ &\  13 \ &\  4\\
        7 \ &\  8 \ &\  9 \ &\  1 \ &\  4 \\
        1 \ &\  9 \ &\  10 \ &\  11 \ &\  4\\
       13 \ &\  1 \ &\  11 \ &\  12 \ &\  4 \\
          \ &\  \ &\  \vdots \ &\  \ &\ \\
         \end{array}\right) \enskip ;\enskip
         \left(\begin{array}{cccc|c}
        2 \ &\  - \ &\  4 \ &\  5 \ &\  3\\
        6 \ &\  - \ &\  1 \ &\  13 \ &\  3\\
        - \ &\  - \ &\  - \ &\  - \ &\  - \\
        1 \ &\  - \ &\  10 \ &\  11 \ &\  3\\
       13 \ &\  1 \ &\  11 \ &\  12 \ &\  4 \\
          \ &\  \ &\  \vdots \ &\  \ &\ \\
         \end{array}\right) \ ;\ 
\left(\begin{array}{cccc|c}
        2 \ &\  - \ &\  - \ &\  5 \ &\  2\\
        - \ &\  7 \ &\  1 \ &\  13 \ &\  3\\
        - \ &\  - \ &\  - \ &\  - \ &\  - \\
        - \ &\  - \ &\  - \ &\  - \ &\  - \\
       13 \ &\  1 \ &\  11 \ &\  12 \ &\  4 \\
          \ &\  \ &\  \vdots \ &\  \ &\ \\
         \end{array}\right)
\]

The first four columns give us the neighbours in these four directions and the last column gives
us the number of neighbours. For the fourth example of a fourfold
coordinated planar random network, the neighbour map is the same for
the square lattice, however the {\bf S} is different. 

The action of {\bf S} on a vector $\vert n\rangle$ is prompted by the
relevant neighbour map. The operations proceed as follows~ :
\begin{enumerate}
\item[(i)] One by one choose  rows  $\ul{n}_i$ of $\vert n\rangle$
\item[(ii)] From the neighbour map  choose $m=N_{ik}$ (which is the $k$-th neighbour
of $i$)
\item[(iii)]  Multiply $\uul{S}_{mi}\ul{n}^T_i$ and  add this to the $m$-th row of a new $\vert n'\rangle$ 
\item[(iv)] Repeat this for all the non-zero rows of $\vert n\rangle$.
 Finally $\vert n'\rangle = {\mathbf S}\vert n\rangle$.
\end{enumerate}

The  the recursive operations described in Eqns. (\ref{rec}) can be encoded into
the following modules : \\

A. \underline{Module HOP}($\vert\Phi_2\rangle ; \vert\Phi_1\rangle$), which describes
the action of the Hamiltonian on a `vector'.\\
\framebox[5.5in]{
\begin{minipage}[t]{5.5in}
\begin{eqnarray*}
\mathrm{The\ input} & \mathrm{into} & \mathrm{this\ module\ is :}
\quad \vert n\rangle \Rightarrow \vert \Phi_2\rangle \\
\phantom{x}& \phantom{x}&\phantom{x} \\
{\bf \Delta}^{1/2}\vert\Phi_2\rangle &\Rightarrow &\vert\Psi_1\rangle\enskip ;\enskip 
{\bf S}\vert\Psi_1\rangle \Rightarrow \vert\Psi_2\rangle \enskip ;\enskip
{\bf \Delta}^{1/2}\vert\Psi_2\rangle \Rightarrow \vert\Psi_1\rangle \\
({\bf C-E}_\nu)\vert \Phi_2 \rangle&\Rightarrow& \vert\Psi_2\rangle\enskip ; \enskip \vert\Phi_1\rangle =
\vert\Psi_2\rangle+\vert\Psi_1\rangle \equiv {\bf h}\vert n\rangle\\
{\bf o}\vert\Phi_1\rangle &\Rightarrow& \vert \Psi_3\rangle \equiv {\bf o}{\bf h} \vert n\rangle\\
{\bf \Delta}^{1/2}\vert\Psi_3\rangle &\Rightarrow &\vert\Psi_1\rangle\enskip ;\enskip 
{\bf S}\vert\Psi_1\rangle \Rightarrow \vert\Psi_2\rangle \enskip ;\enskip
{\bf \Delta}^{1/2}\vert\Psi_2\rangle \Rightarrow \vert\Psi_1\rangle \\
({\bf C-E}_\nu)\vert \Psi_3 \rangle&\Rightarrow& \vert\Psi_2\rangle\enskip ; \enskip \vert\Phi_1\rangle =
\vert\Phi_1\rangle - \vert\Psi_2\rangle-\vert\Psi_1\rangle \equiv \left[{\bf h}-{\bf hoh} \right]\vert n\rangle\\
\phantom{x}& \phantom{x}&\phantom{x} \\
{\bf E}_\nu\vert\Phi_2\rangle &\Rightarrow &\vert\Psi_1\rangle \enskip ;\enskip \vert\Phi_1\rangle=
\vert\Phi_1\rangle+\vert\Psi_1\rangle \equiv {\bf H}^{(2)}\vert n\rangle \\
\phantom{x}& \phantom{x}&\phantom{x} \\
\mathrm{The\ output}&\mathrm{from}& \mathrm{this\ module\ is}\quad
\vert\Phi_1\rangle\Rightarrow {\bf H}^{(2)}\vert n\rangle
 \end{eqnarray*}
\end{minipage}
}
B. The \underline{Module REC}( $\vert\Phi_2\rangle,\vert\Phi_3\rangle,\beta_n^2; \alpha_n,\beta_{n+1}^2$)
which calculates the coefficients $\alpha_n,\beta^2_{n+1}$ recursively :\\
\framebox[5.5in]{
\begin{minipage}[t]{5.5in}
\begin{eqnarray*}
\mathrm{The\ input\ }&\mathrm{into\ }&\mathrm{this\ module\ is\ :}\quad \vert\Phi_2\rangle \equiv \vert n\rangle \ ;\ \vert\Phi_3\rangle\equiv \vert n-1\rangle\ \mbox{and}\ \beta^2_n \\
\phantom{x}& \phantom{x}&\phantom{x} \\
\mathrm{Run\ }&\Rightarrow& \mathrm{HOP}(\vert\Phi_2\rangle;\vert\Phi_1\rangle)\quad \alpha_n =\langle \Phi_1\vert\Phi_1\rangle/\langle \Phi_2\vert\Phi_2\rangle\\ 
\vert\Phi_1\rangle & = & \vert\Phi_1\rangle -\alpha_n*\vert\Phi_2\rangle -\beta^2_n*\vert\Phi_3\rangle\quad
\beta^2_{n} = \langle\Phi_1\vert\Phi_1\rangle/\langle \Phi_2\vert\Phi_2\rangle\\
\vert\Phi_3\rangle & = & \vert\Phi_2\rangle \mbox{ and } \vert\Phi_2\rangle = \vert\Phi_1\rangle \\
\phantom{x}& \phantom{x}&\phantom{x} \\
\mathrm{The\ output }&\mathrm{from }&\mathrm{this\ module\ is:}\ \vert\Phi_2\rangle \equiv \vert n+1\rangle  ; \vert\Phi_3\rangle\equiv \vert n\rangle\ \mbox{and}\ \alpha_n,\beta^2_{n} \\
\end{eqnarray*}
\end{minipage}
}

The two vectors $\vert\Phi_1\rangle$ and $\vert\Phi_2\rangle$  are input/output  vectors
of the Module HOP, while the three vectors  $\vert\Psi_1\rangle,\vert\Psi_2\rangle$ and $\vert\Psi_3\rangle$ are dummies which are needed within the module. The space for these
dummies may be dynamically allocated during this procedure and released after the operation is over.
The same is true for the vector $\vert\Phi_1\rangle$ in the Module REC. \\

Eqn. (\ref{albe}) indicates that in the new basis the Hamiltonian
representation is tri-diagonal. It follows immediately that The Green function
for the system is given by a continued fraction :

\begin{eqnarray} 
G_{\R_iL,\R_iL}(z)  =  \langle 1\vert G(z)\vert 1\rangle 
 =  \frac{\displaystyle 1}{\displaystyle z-\alpha_1-
\frac{\displaystyle \beta^2_1}{\displaystyle z-\alpha_2-\frac{\displaystyle \beta^2_2}
{\displaystyle \ddots  
z-\alpha_{n_0}-\beta^2_{n_0}T(z)}}}\nonumber \\
\label{gf}
\end{eqnarray}

The asymptotic part of the Green function $T(z)$ is called the {\sl Terminator}. Many terminators
are available in literature. The suitable one depends upon the way in which the coefficients
$\alpha_n,\beta^2_n$ behave as $n\rightarrow\infty$. The most commonly used is the square root
terminator which is suitable when $\{\alpha_n,\beta^2_n\}\rightarrow\{\alpha,\beta^2\}$. Luchini and Nex \cite{ln} have suggested a modification
when calculations up to a large $n_0$ is not possible.
If we carry out the terminator approximation after $n_0$ steps the first $2n_0$ moments of the
density of states are {\it exact} and it has been shown that the asymptotic moments are also
accurately reproduced. Beer and Pettifor \cite{bp} have suggested an alternative way of
obtaining the terminator. They note that $T(z)=T(z,\{\alpha_n,\beta_n\},n\leq n_0)$ so that
the Beer-Pettifor prescription is a  closed algorithm. It wins over the square-root terminator
described above when the convergence of the coefficients is either oscillatory or slow. Viswanathan
and Muller \cite{vm} have suggested several other terminators like the terminator with an exponential
 tail, suitable when $\beta^2_n\rightarrow n$ and the terminator with a Gaussian  tail, suitable
when $\beta^2_n\rightarrow n^2$.  We shall have
options for several terminating procedures so that the user may choose according to his need.
C. The \underline{Module GREEN}($L_{max},n_0,E_{min},E_{max}$) which calculates the Green function.\\
\framebox[5.5in]{
\begin{minipage}[t]{5.5in}
\begin{eqnarray*}
\mathrm{Loop}& & (L=1,L_{max})  \\
\mathrm{Input} &\Rightarrow & \vert\Phi_2\rangle = \vert\R_iL\rangle,\ \vert\Phi_3\rangle =\vert 0\rangle\ ;\ \beta^2_0=1\\
\quad \mathrm{Loop} & & (n=1,n_0) \\
\mathrm{Run} & \Rightarrow & \mathrm{REC}(\vert\Phi_2\rangle;\vert\Phi_3\rangle;\alpha_n,\beta^2_n) \\
& & \mathrm{End\ Loop}\\
 & & \mathrm{End\ Loop} \\
\mathrm{Loop}& & (L=1,L_{max})  \\
\mathrm{Loop} & & (E=E_{min},E_{max}) \\
\mathrm{Run} &\Rightarrow & \mathrm{TERM}(\{\alpha_n,\beta^2_n\},n_0, E, G) \\
\mathrm{Loop} & & (n=n_0,1,-1) \\
G & = & \frac{1}{E-\alpha_n-\beta^2_n*G}\\
& &  \mathrm{End\ Loop}\\
 & & \mathrm{End\ Loop} \enskip\Rightarrow\enskip\mbox{Output} = G_{\R_iL,\R_iL}(E) 
\end{eqnarray*}\end{minipage}}

The partial
density of states, projected onto a particular $(\R_iL)$, is :

\be
 n_{\R_iL}(E) = -\frac{1}{\pi}\ \lim_{\delta\rightarrow 0} \Im m \  
G_{\R_iL,\R_iL}(E+i\delta) \ee

It is clear from this discussion that the recursion procedure is general enough to deal with lattices
of any complexity, surfaces and interfaces and disordered networks.

\section{The augmented space formalism}
Finally we come to the main part of the package : that which deals with disorder.
The augmented space formalism \cite{ast} allows us to directly compute the configuration average of 
the Green function by constructing a Hamiltonian in the augmented space of configurations of the random
parameters. The formalism is based on ideas prevalent in quantum
measurement theory : we associate with each random parameter $n_i$ an operator $N_i$ whose spectrum 
 $\{n^\lambda_i\}$ are  the measured values of the parameter. The configuration state of $n_i$ are
the eigenkets $\{\vert n^\lambda_i\rangle\}$ of $N_i$, which, therefore is an operator in the
space ${\cal{C}}_i$ spanned by the different configuration states. Given this, the probability density of the parameter is

\[ p(n_i) = -\frac{1}{\pi}\ \Im m  \ \langle \emptyset_i \vert\left(\rule{0mm}{3mm} (n_i+i0)I - N_i\right)^{-1}
\vert \emptyset_i\rangle \]

where 

\[ \vert\emptyset_i\rangle = \sum_\lambda\ \sqrt{p(\lambda)} \vert n^\lambda_i\rangle \]

The augmented space theorem \cite{ast} then gives us the configuration average of {\sl any} function of the random variables~:

\be \ll \Phi(\{n_i\})\gg = \langle \emptyset \vert \widetilde{\Phi}(\{N_i\})\vert \emptyset \rangle \ee

Here,
\begin{enumerate}
\item[(i)] $\vert\emptyset\rangle = \prod^\otimes_i\  \vert\emptyset_i\rangle $. It is a member of the basis
set $\vert\{C\}\rangle\ =\ \{\vert n_1^{\lambda_1},n_2^{\lambda_2},\ldots \rangle\}$ which spans the configuration space $\Psi$ of the set of random parameters $\{n_i\}$.
\item[(ii)] $\widetilde{\Phi}(\{N_i\})$ is an operator in the configuration space $\Psi=\prod^\otimes {\cal{C}}_i$. It is the
{\sl same} operator functional of $\{N_i\}$ as $\Phi(\{n_i\})$ was a function of $\{n_i\}$.
\item[(iii)] The expression is exact. The configuration average is done exactly first, then the approximations
may be carried out maintaining various physical constraints. The philosophy is very different from the
mean-field approximations like the CPA.
\end{enumerate}

If we apply this to the configuration averaged Green function of a random
Hamiltonian $H^{(2)}(\{n_i\})$ :

\begin{eqnarray} \ll G_{\R_iL,\R_iL}(z,\{n_i\})\gg =
\langle R_i\otimes\emptyset\vert \left(\rule{0mm}{3mm}
z\ul{\ul{\widetilde{I}}}-\ul{\ul{\widetilde{H}}}^{(2)}(\{\widetilde{N}_i\}\right)^{-1}\vert R_i\otimes\emptyset\rangle\nonumber\\
 \label{avgreen}\end{eqnarray}

Let us take as an example the case of a disordered binary alloy : the Hamiltonian has the same form as Eqn. (\ref{ham}) but~:

\begin{eqnarray*}
\ul{\ul{\widehat{C}}}_i  &=& \ul{\ul{\widehat{C}}}_A\ n_i
 + \ul{\ul{\widehat{C}}}_B \ (1-n_i)\quad\mbox{where}\quad
\ul{\ul{\widehat{C}}}_i = \ul{\ul{C}}_i - \ul{\ul{E}}_\nu \quad\mbox{and}\quad\\
\ul{\ul{\Delta}}^{1/2}_i & =&  \ul{\ul{\Delta}}^{1/2}_A\ n_i+\ul{\ul{\Delta}}^{1/2}_B\ (1-n_i)
\quad\mbox{and}\quad
\uul{o}_i  =  \ul{\ul{o}}_A\ n_i+\ul{\ul{o}}_B\ (1-n_i)
\end{eqnarray*}

The random ``occupation" variables  $n_i$ take the values 0 and 1 with probabilities proportional
of the concentrations x and y of the constituents A and B.  $N_i$ has a $2\times 2$ representation :
 $\left(\begin{array}{cc}
                                                                x & \sqrt{xy}\\
                                                               \sqrt{xy}& y
							       \end{array}\right)$
. If we denote the basis of the above representation by $\vert\uparrow_i\rangle, \vert\downarrow_i\rangle$, then 
\[ \widetilde{N}_i = x {\cal I} + (y-x){\cal P}_{\downarrow_i} + \sqrt{xy} {\cal T}_{\uparrow_i\downarrow_i}
\]

and each of the potential parameters become operators in the configuration space, for example : 

\begin{eqnarray}
\widetilde{\mathbf C} &=& \sum_i \left\{\rule{0mm}{3mm} \ll \uul{\hat{C}}_i\gg \ {\cal I} + \ B(\uul{\hat{C}}_i) \ {\cal P}_{\downarrow_i} + \ F(\uul{\hat{C}}_i) \ {\cal T}_{\uparrow_i\downarrow_i}\right\}\otimes {\cal P}_i
\nonumber\\
{\widetilde{\mathbf\Delta}}^{1/2} & = & \sum_i \left\{\rule{0mm}{3mm} \ll \uul{{\Delta}}^{1/2}_i\gg \ {\cal I} + \ B(\uul{{\Delta}}^{1/2}_i) \ {\cal P}_{\downarrow_i}
+ \ F(\uul{{\Delta}}^{1/2}_i) \ {\cal T}_{\uparrow_i\downarrow_i}\right)\otimes {\cal P}_i
\nonumber\\
{\widetilde{\mathbf E}}_\nu &=& \sum_i \left\{\rule{0mm}{3mm} \ll \uul{{E}}_\nu\gg \ {\cal I} + \ B(\uul{{E}}_\nu) \ {\cal P}_{\downarrow_i} + \ F(\uul{E}_\nu) \ {\cal T}_{\uparrow_i\downarrow_i}\right\}\otimes {\cal P}_i
\nonumber\\
{\widetilde{\mathbf o}}& = &\sum_i \left\{\rule{0mm}{3mm} \ll \uul{{o}}_i\gg \ {\cal I} + \ B(\uul{{o}}_i) \ {\cal P}_{\downarrow_i}
+ \ F(\uul{\hat{o}}_i) \ {\cal T}_{\uparrow_i\downarrow_i}\right\}\otimes {\cal P}_i
\label{aug}
\end{eqnarray}

where $B(C) = (y-x)(C_A-C_B)$ and $F(C) = \sqrt{xy}(C_A-C_B)$. Also, since the structure matrix is not random :

\be {\widetilde{\mathbf S}} = \sum_{ij} \uul{S}_{\R_i\R_j}\ {\cal I}\otimes {\cal T}_{ij} \label{str}\ee

\begin{eqnarray}
{\widetilde{\mathbf H}}^{(2)} & =&   {\widetilde{\mathbf E}}_\nu + {\widetilde{\mathbf h}}- 
{\widetilde{\mathbf h}}\ {\widetilde{\mathbf o}}\ {\widetilde{\mathbf h}}\quad\mbox{where}\quad
{\widetilde{\mathbf h}}  =   {\widetilde{\mathbf C}} + {\widetilde{\mathbf \Delta}}^{1/2} \otimes
{\widetilde{\mathbf S}}  \otimes {\widetilde{\mathbf \Delta}}^{1/2}
\label{ham-as}
\end{eqnarray}

If we now compare Eqns. (\ref{avgreen}) and (\ref{ham-as}) with Eqns. (\ref{ham2})-(\ref{gf}) it becomes clear that
once we have defined the Hamiltonian in the  space of configurations, the recursion method may be directly used
to obtain the configuration averaged Green function. The approximation involved will then only be the
``termination" approximation. Heine's ``Black-body theorem" \cite{vol35} indicates that most electronic structure
energetics is dominated by the immediate environment of a site. This is a major justification of the termination
approximation. Unlike the CPA which gives only the first eight moments of the density of states exactly and
the asymptotic moments accurately, the augmented space recursion gives $2n_0$ moments exactly (where the termination
is done after $n_0$ steps) and the asymptotic moments also accurately. In most of our calculations we can
take $n_0$ about 8-9 steps.

\section{The TB-LMTO-ASR Algorithm}

The full package is divided into
several modules. The first  is the Preparation Module. This module has two
branches, one each for each constituent of the alloy. Each branch runs
parallely in two different slave processors. The structure matrix is
prepared in the master processor. This Module prepares the following :
\begin{enumerate}
\item[(i)] It prepares the two control files for the different alloy constituents.
This part is interactive with the user.
\item[(ii)] It carries out the simple Hartree calculation and prepares the atomic
radii of the constituents and the initial charge density.
\item[(iii)] It calculates the overlap between atomic spheres and inputs empty
spheres maintaining the symmetries.
\item[(iv)] It calculates the structure matrix from the inputs of the control files.
\end{enumerate}

\begin{figure} 
\centering
\includegraphics[width=6in,height=7in]{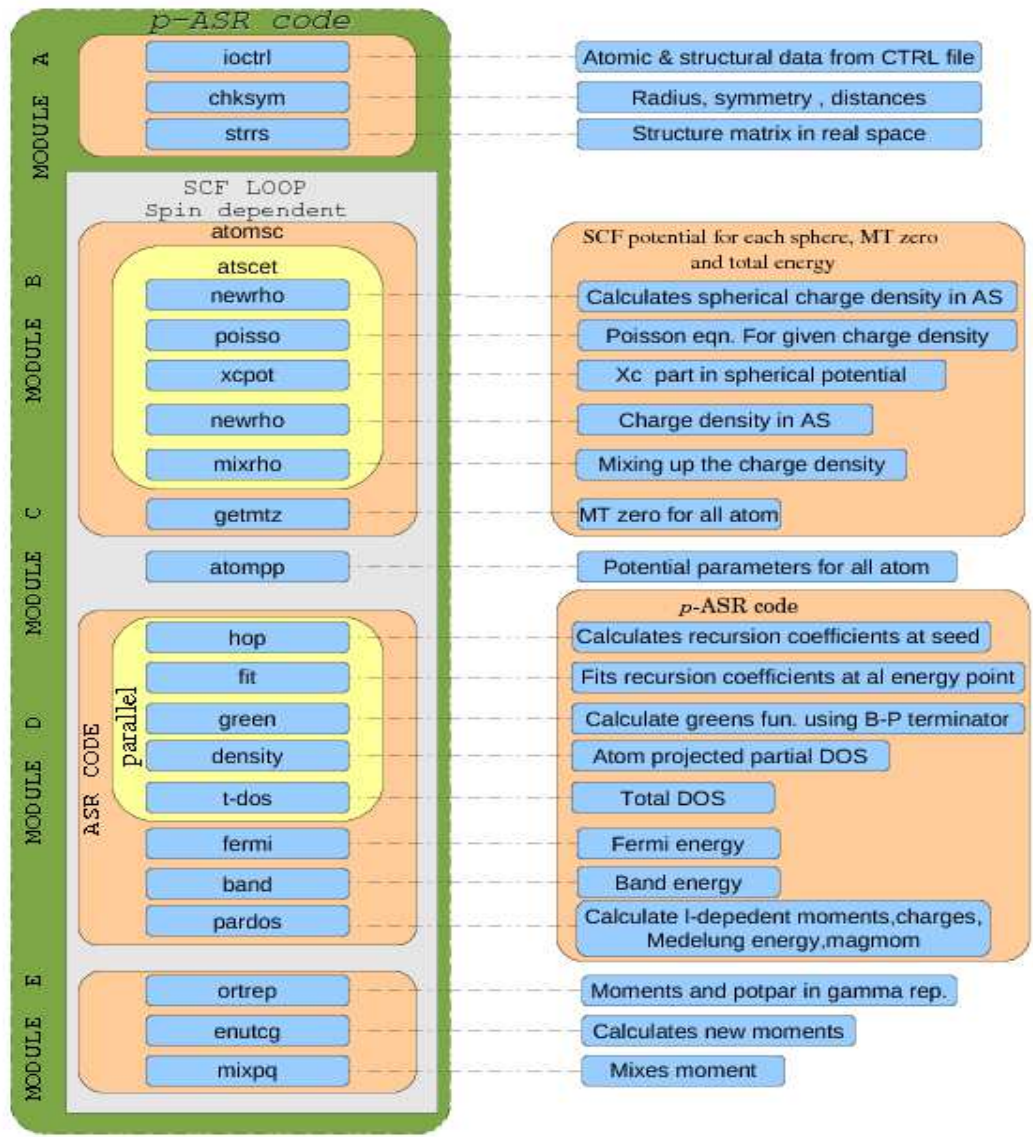}
\caption{(Color Online) The TB-LMTO-Augmented Space Recursion package flowchart}\label{pflow}
\end{figure}

The Preparation Module is followed by the main ASR-module. This module is called by the
routine {\sl lmasr}.  This main module is divided into five smaller  modules :

\begin{itemize}
\item[(i)] \underline{Module A} : This module reads the data generated by the Preparation
Module. It checks the inputs for consistency. The routines in this module are again those
in the Stuttgart LMTO47, but modified to read the inputs for {\sl both} the two constituents
of the alloy.
\vskip 0.4cm
\item[(ii)] \underline{Module B}: At the start of the DFT self-consistency loop, this module takes
the overlap of a simple Hartree atomic density calculation in the Preparation module and generates
the Hartree and exchange-correlation potentials, spheridizes them and inputs them to the Atomic Module
that follows. In later steps of the self-consistency loop, this Module first mixes the charge
densities of the earlier steps and prepares the Hartree and exchange-correlation potentials,spheridizes 
them for input into the Atomic Module. 
\vskip 0.2cm

   - At this point there is a choice of using either the standard DFT exchange-correlation potentials or, alternatively,
there is a branch module {\sl Harbola-Sahni} which sets up the Harbola-Sahni potential for the study of
excited states \cite{rgsmh}.
\vskip 0.4cm

\item[(iii)] \underline{Module C}: This is the Atomic Module. It takes the spheridized 
Kohn-Sham potential generated in Module B and solves the radial Kohn-Sham equation numerically.
The Kohn-Sham orbitals and energies then lead to the potential parameters for each constituent.
Those for the two constituents are calculated on different processors. The parameters are 
first calculated in the orthogonal representation. A new routine {\sl gtoa}, not present in the 
Stuttgart LMTO47 package, then transforms them to the most screened tight-binding representation.
\vskip 0.4cm

\item[(iv)] \underline{Module D} : This is the main ASR Module. The routines herein have been
fully developed by us and form the main backbone of the package. The input are the tight-binding
Hamiltonian parameters from the Atomic Module. First they are combined with the alloy composition
to prepare the augmented space Hamiltonian. The nearest neighbour map in augmented space is then
generated. Next, the recursion is carried out for each $L$ value, terminators generated and the
$L$ projected density of states are calculated. Each different recursion for each $L$
value is carried out on a different processor, thus vastly accelerating the calculations.

We then proceed to calculate the total density of states and the Fermi energy. Again, branching
out into different processors, we calculate the $L$-dependent moments and magnetic moments.
 Of all the Modules, this is the one amenable to maximum parallelization.
\vskip 0.2cm

 - At this point we have the possibility of introducing short ranged order.The ASR for short-ranged
order has been described in some detail earlier \cite{prasad}.  The branching for this
choice occurs just before we set up  the augmented space Hamiltonian. The extra input is the
Warren-Cowley short ranged order parameter.
\vskip 0.2cm

 - Also at this point we have the option to introducing disorder in the structure matrix because of
size mismatch between the two constituents of the alloy. The branching now takes place earlier
in the Preparation Module where we generate not one, but three different structure matrices :
$\uul{S}^{AA}_{ij}, \uul{S}^{BB}_{ij}$ and $\uul{S}^{AB}_{ij}$. In the `end point' approximation 
 \cite{sm} the Eqn. (\ref{str}) is now replaced by :

\be {\widetilde{\mathbf S}} = \sum_{ij} \left\{\rule{0mm}{3mm}\ll\uul{S}_{ij}\gg {\cal I}
+ \uul{S}^{(1)}_{ij} \left({\cal N}_i+{\cal N}_j\right) 
+ \uul{S}^{(2)}_{ij} {\cal N}_i\otimes {\cal N}_j \right\} \otimes{\cal T}_{ij}
\ee

\n where ${\cal N}_i = (y-x){\cal P}_{\downarrow_i} + \sqrt{xy} {\cal T}_{\uparrow_i\downarrow_i}$,\quad  $\uul{S}^{(1)}=
\uul{S}^{AB}_{ij}-\uul{S}^{BB}_{ij}$ and  $\uul{S}^{(2)}= \uul{S}^{AA}_{ij}+\uul{S}^{BB}_{ij}- 2\uul{S}^{AB}_{ij}$. 
This modification will be  available in this module.
\vskip 0.2cm

It is in these last two options that the ASR really scores over the CPA, which cannot really deal with
either short-ranged order or off-diagonal disorder as both involve more than one site. The competing methodology
is the special quasi-random structures (SQS) \cite{sqs}. However, if we are dealing with materials with many
atoms per unit cell and non-stoichiometric compositions, the SQS required will be rather large and  will involve
use of {\sl huge} unit cells. Here the TB-LMTO-ASR with the use of much smaller unit cells will score.
 We have shown earlier \cite{fecr} that both these two methods give virtually the same results for the density of states.
\vskip 0.4cm

\item[(v)] \underline{Module E} : In this module the $L$-dependent moments are used to obtain the charge
density. We also calculate the total energy, including the Madelung term. For the disordered alloy, 
the Madelung term is obtained from the procedure suggested by Skriver and Ruban \cite{mad}.
Finally the old and new moments are mixed, the mixed charge density thus obtained  is input
back in Module B. This  is iterated till convergence in both energy and charge density is achieved.
\end{itemize}

\begin{figure}[h!] 
\centering
\vskip 1cm
\includegraphics[scale=.5]{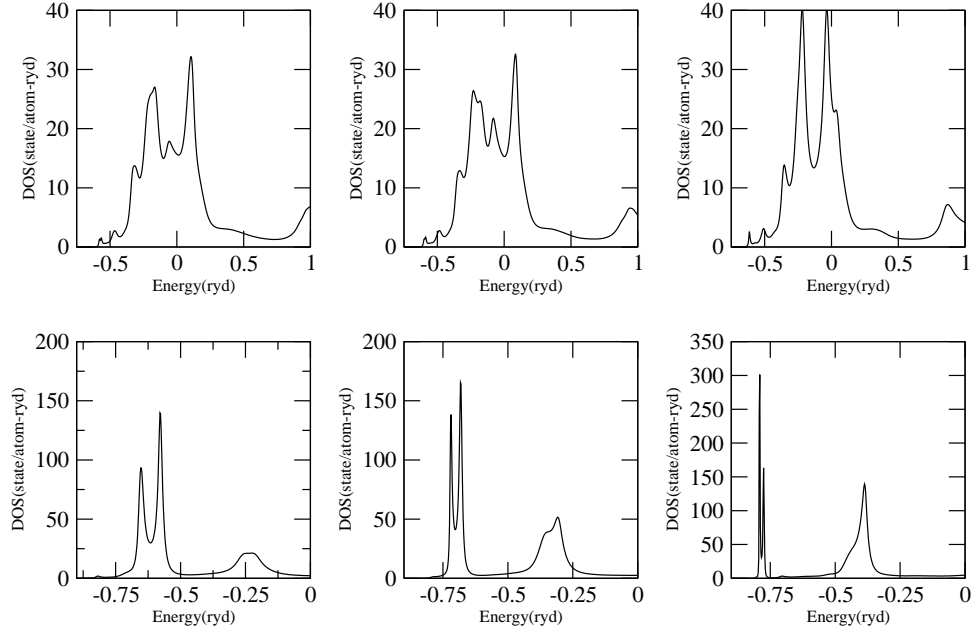} 
\caption{	(Top) DOS for Cr$_x$Fe$_{1-x}$ for different compositions : 
(left) Cr$_{7}$Fe$_{3}$;  (middle)  Cr$_{5}$Fe$_{5}$; (right) Cr$_{3}$Fe$_{7}$; 
(Bottom) DOS for Cu$_x$Zn$_{1-x}$ for different compositions : (left) Cu$_{24}$Zn$_{76}$; (middle) Cu$_{50}$Zn$_{50}$; (right) Cu$_{75}$Zn$_{25}$.
Energy(Ryd) is plotted along $ x $-axis; $ y $-axis is DOS (states/Ryd)}
\label{dos}
\end{figure}

For the DFT part, our code depends heavily on Stuttgart-LMTO routines developed by  Anderson and co-workers \cite{aj}
 Two independent DFT codes run parallely to produce the potential parameters of the two constituent atoms
of the binary alloy. These potential parameters are used at the input point of our ASR routines in Module D. 

Our present DFT modules deal only with collinear magnetism. For spin dependent calculations the Hamiltonian
is separable in spin space. Thus the spin is merged with the label $L$ which is now $\{\ell,m,\sigma\}$ and
we have just to carry out twice the number of recursions : for $\sigma = \uparrow$ and $\downarrow$. In case
we wish to introduce spin-orbit coupling and possibility of non-collinear magnetism, we have to replace
the DFT module with one dealing with density matrices, rather than densities \cite{noncol} and the ASR
module with one applying generalized or vector recursion \cite{genrec1}-\cite{genrec2}  which was designed to deal with
 Hamiltonians  whose representation in spin-space or `spinor' bases is not diagonal :

\[ \ll \uul{n}\vert \uul{\widetilde{H}}\vert \uul{m}\gg = \left( \begin{array}{cc}
 \uul{H}^{\uparrow\uparrow}_{nm} & \uul{H}^{\uparrow\downarrow}_{nm}\\
 \uul{H}^{\downarrow\uparrow}_{nm} & \uul{H}^{\downarrow\downarrow}_{nm}
\end{array}\right) 
 \enskip\mbox{where }\enskip   \vert \uul{n}\gg = \left(\begin{array}{c}
  \vert n\!\!\uparrow\rangle\\
  \vert n\!\!\downarrow\rangle
  \end{array}\right) \]

This new module is under preparation \cite{gam} and will be incorporated once the basic checks are carried out. Here
we mention this in order to bring out the different possibilities and versatility of the package.

\section{Applications and performance analysis}

We have described the contents and commented upon the efficiency and versatility of the
TB-LMTO-ASR package. We shall conclude by describing two different applications : namely,
the alloys $\textrm{Cu}_x\textrm{Zn}_{1-x}$ where both the constituents are non-magnetic
in the bulk and the $d$-band centres of Cu and Zn are well separated;  and Fe$_x$Cr$_{1-x}$
where both constituents are magnetic and their $d$-bands overlap.

\begin{table}[ht]
\begin{center}
\begin{tabular}{lcc} \hline
\multicolumn{3}{c}{For a single self-consistency loop}\\ \hline
& Old TB-LMTO-ASR & New TB-LMTO-ASR \\
Wall time & 699 sec & 225 sec \\
Efficiency$\left( \frac{\rm cputime}{\rm walltime}\right) $ & 0.714  & 0.97 \\
\hline
\end{tabular}
\end{center}
\label{comp}
\end{table}

The alloys have been studied earlier by us using the old version of the TB-LMTO-ASR \cite{asr2},\cite{oasr1}-\cite{oasr2} and TB-LMTO-CCPA\footnote{Tight-binding linear muffin-tin orbitals cluster coherent potential approximation} \cite{ccpa}, both introduced
by us, and the KKR-ICPA\footnote{Korringa-Kohn-Rostocker itinerant coherent potential approximation} \cite{icpa}
 which is also based on the augmented space formalism, as well as through the KKR-CPA\footnote{Korringa-Kohn-Rostocker coherent potential approximation} \cite{kkrcpa}, PAW-SQS\footnote{Projector augmented wave special quasi-random structures}\cite{sqs} and KKR-NL-CPA\footnote{Korringa-Kohn-Rostocker non-local coherent potential approximation} \cite{nlcpa1}-\cite{nlcpa2}. They are therefore
ideal system for benchmarking the new TB-LMTO-ASR package. Comparison of the densities of states shown in 
Fig. \ref{dos} with the results shown in the above references will convince us that for Fe$_x$Cr$_{1-x}$ 
there is hardly anything
to choose between the various techniques and the packages based on them. However, for the split band alloy Cu$_x$Zn$_{1-x}$ the TB-LMTO-ASR, TB-LMTO-CCPA and KKR-NL-CPA scores over the CPA versions, as expected from earlier analysis.

One of our our main point of interest is the relative runtime and efficiency of this new version of the TB-LMTO-ASR, as compared 
to the several earlier versions proposed by us. To benchmark these characteristics we take a specific calculation
on Fe$_x$Cr$_{1-x}$.  
Each calculation is done on a  73109 site augmented space map, with N recursion steps with $L=s,p_x,e_g$ and $t_{2g})$,
and $\sigma = 1,2$, followed  by a Beer-Pettifor termination scheme . 
The old and new versions of the TB-LMTO-ASR are compared in the Table \ref{comp}. 
Here Wall time is the same  as run time and Efficiency is the ratio between the Wall time and CPU time. 
Gnu profiles~(gprof)\footnote{http://www.cs.utah.edu/dept/old/texinfo/as/gprof\_toc.html}  for the new TB-LMTO-ASR
 output and the old serial version are given in the Table (\ref{gpr}). 

\begin{table}[h]
\begin{center}
\begin{scriptsize}\begin{tabular}{|c|c|c|c|c|c|c|}\hline
\%  & cumulative&   self  & &           self  &   total&\\
time &  seconds&   seconds&    calls &  s/call&   s/call & name\\\hline
49.91 (61.47)  &   100.73 (397.33) &  100.73 (397.33) &      81 (128)   &  1.24 (3.10) &    2.26 (5.01)& hop \\
 40.78 (37.67) &   183.04 (640.87) &   82.31 (243.54) &1.6$ (3.9) \times 10^9$&  0.00 (0.00) &    0.00 (0.00) & matp \\
  0.76 (0.79) &   196.82  (646.41)&    1.53 (5.11) &       7(16)   &  0.22 (0.00) &    0.22 (0.32)& spectral\\
  0.00 (0.00)&   201.84 (646.40) &    0.00 (0.01) &  1.8 (3.5) $\times 10^5$  &  0.00 (0.00)&    0.00(0.00)& splint  \\
  0.00 (0.00) &   201.84 (646.44) &    0.00 (0.00) &     206 (384)   &  0.00 (0.00) &    0.00 (0.00) & spline \\
  0.00 (0.00) &   201.84 (646.44) &    0.00 (0.00) &      56 (56)   &  0.00 (0.00) &    0.00 (0.00) & matmult \\
  0.00 (0.00) &   201.84 (646.44) &    0.00 (0.00) &      36  (36)  &  0.00 (0.00) &    0.00 (0.00)& mom \\
  0.00(*)  &   201.84 (*) &    0.00 (*) &      28 (*) &  0.00 (*) &    6.59 (*) & doparallel \\
  0.00 (0.00) &   201.84 (646.44) &    0.00 (0.00)&       7 (16)  &  0.00 (0.00)&    0.00 (0.00) & fit \\
  0.00 (0.00) &   201.84 (646.44)&    0.00 (0.00)&       4 (7) &  0.00 (0.00) &    0.00 (0.00) & tdos \\
  0.00 (0.00) &   201.84 (646.44) &    0.00 (0.00) &       1 (1)  &  0.00 (0.00) &    0.00 (0.00) & band \\
  0.00 (0.00) &   201.84 (646.44) &    0.00 (0.00) &       1 (1)   &  0.00 (0.00) &    0.00 (0.00) & fermi \\
  0.00 (0.00)&   201.84 (646.44) &    0.00 (0.00)&       1 (1)   &  0.00 (0.00)&    0.00 (0.00) & pardos \\\hline
\end{tabular}\end{scriptsize}
\end{center}
\begin{center}
\caption{gprof data for the optimized and parallelized ASR code and (in brakets) the old serial code. The discontinuity between the cumulative time of \texttt{spectral} and \texttt{splint} is due to the machine routine which are not included here. }\label{gpr}
\end{center}
\end{table}

\section{Conclusion}
We have presented here a computational package that combines three different techniques to allow
us to study the electronic properties of disordered systems. The package can deal with bulk
disordered alloys as well as surfaces and interfaces, both smooth, stepped  and rough and also
structurally distorted lattices. The package can be generalized at many stages which have  been
clearly commented upon and the generalizations are being carried out step by step.


\end{document}